\documentclass[usenatbib]{mn2e}
\usepackage{graphicx}
\usepackage{lscape}
\usepackage{amssymb}
\usepackage{amsmath}
\usepackage{url}
\usepackage{aas_macros}
\usepackage{multirow}

\newcommand{\kmspc}{km\,s$^{-1}$\,pc$^{-1}$}
\newcommand{\kms}{km\,s$^{-1}$}

\title[Age of the BPMG]{On the age of the $\beta$ Pictoris moving group}

\author[Mamajek \& Bell]{
Eric E. Mamajek\thanks{Email: emamajek@pas.rochester.edu} 
and 
Cameron P. M. Bell\\
Department of Physics \& Astronomy, University of Rochester, Rochester, NY 14627, USA}

\begin{document}

\date{Accepted ?, Received ?; in original form 2014 July 29}

\pagerange{\pageref{firstpage}--\pageref{lastpage}} \pubyear{2014}

\maketitle

\label{firstpage}

\begin{abstract}
Jeffries \& Binks (2014) and Malo et al. (2014) have recently reported
Li depletion boundary (LDB) ages for the $\beta$ Pictoris moving group
(BPMG) which are twice as old as the oft-cited kinematic age of $\sim
12\,\rm{Myr}$.
In this study we present (1) a new evaluation of the internal
kinematics of the BPMG using the revised \emph{Hipparcos} astrometry
and best available published radial velocities, and assess whether a
useful kinematic age can be derived, and (2) derive an isochronal age
based on the placement of the A-, F- and G-type stars in the
colour-magnitude diagram (CMD).
We explore the kinematics of the BPMG looking at velocity trends along
Galactic axes, and conducting traceback analyses assuming linear
trajectories, epicyclic orbit approximation, and orbit integration
using a realistic gravitational potential.
None of the methodologies yield a kinematic age with small
uncertainties using modern velocity data.
Expansion in the Galactic $X$ and $Y$ directions is significant only
at the 1.7$\sigma$ and 2.7$\sigma$ levels, and together yields an
overall kinematic age with a wide range ($13-58\,\rm{Myr}$; 95 per
cent CL).
The A-type members are all on the zero age-main-sequence, suggestive
of an age of $> 20\,\rm{Myr}$, and the loci of the CMD positions for the
late-F- and G-type pre-main-sequence BPMG members have a median
isochronal age of 22\,Myr ($\pm$\,3\,Myr stat., $\pm$\,1\,Myr sys.) when
considering four sets of modern theoretical isochrones.
The results from recent LDB and isochronal age analyses are now in
agreement with a median BPMG age of 23\,$\pm$\,3 Myr (overall
1$\sigma$ uncertainty, including $\pm 2\,\rm{Myr}$ statistical and
$\pm 2\,\rm{Myr}$ systematic uncertainties).
\end{abstract}


\section{Introduction
\label{introduction}}

The $\beta$ Pictoris moving group (hereafter BPMG) is a kinematic
group of dozens of young stars in the solar neighbourhood
\citep{Zuckerman04, Torres08}. Such moving groups represent the bridge
between the youngest star-forming regions and the older field star
population, thereby providing us with crucial snapshots of stellar
evolution and physical processes occurring at intermediate ages. The
BPMG is unique in its combination of youth and proximity, and these
factors have lent it to be one of the few stellar groups where the
\emph{Hipparcos} mission \citep{Perryman97} was able to measure
precise astrometry for large numbers of individual members. Its close
distance and young age also makes the BPMG an ideal candidate for
discovering extrasolar planets via direct imaging
(e.g. \citealp{Biller13,Males14}) as well as studying resolved debris
disks with high angular resolution observations
(e.g. \citealp{Boccaletti09}; \citealp*{Churcher11}) and identifying
substellar companions which offer crucial testbeds for evolutionary
models (e.g. \citealp{Mugrauer10,Jenkins12}). Given the group's
importance, it is therefore perturbing that the literature ages for
this benchmark population range from $\sim10-40\,\rm{Myr}$. In
Table~\ref{tab:BPMG_ages} we list the literature age estimates for the
BPMG (see also \citealp*{Fernandez08}).

\begin{table*}
\caption[]{Literature age estimates for the BPMG. We adopt the terms ``traceback age" and ``expansion age" generically for any age estimate trying to infer when an unbound group of stars was at its minimum size in the past.}
\centering
\begin{tabular}{l c l}
\hline
Reference&Age&Method\\
...&(Myr)&...\\
\hline
\protect\cite{Barrado99}  &$20\,\pm\,10\,\rm{Myr}$  &CMD isochronal age (KM stars)\\
\protect\cite{Zuckerman01}&$12^{+8}_{-4}\,\rm{Myr}$&H-R diagram isochronal age (GKM stars) + Li depletion\\
\protect\cite{Ortega02}   &$11.5\,\rm{Myr}$       &Traceback age\\
\protect\cite{Song03}     &$12\,\rm{Myr}$         &Traceback age\\
\protect\cite{Ortega04}   &$10.8\,\pm\,0.3\,\rm{Myr}$&Traceback age\\
\protect\cite{Torres06}   &$\sim18\,\rm{Myr}$     &Expansion age\\
\protect\cite{Makarov07}  &$22\,\pm\,12\,\rm{Myr}$&Traceback age\\
\protect\cite{Mentuch08}  &$21\,\pm\,9\,\rm{Myr}$ &Li depletion\\
\protect\cite{Macdonald10}&$\sim40\,\rm{Myr}$     &Li depletion (magneto-convection models)\\
\protect\cite{Binks14}    &$21\,\pm\,4\,\rm{Myr}$ &Li depletion boundary\\
\protect\cite{Malo14b}     &$26\,\pm\,3\,\rm{Myr}$ &Li depletion boundary\\
\protect\cite{Malo14b}     &$21.5\,\pm\,6.5\,\rm{Myr}$ ($15-28\,\rm{Myr}$) &H-R diagram isochronal age (KM stars)\\
\hline
This work                 &$22\pm3\,\rm{Myr}$     &CMD isochronal age (FG stars)\\
\hline
\multirow{2}{*}{\bf Final}               &{\bf 23 $\pm$ 3 Myr (1$\sigma$)}   & Li depletion bounday \&\\
                          &{\bf [$\pm$2 Myr (stat.), $\pm$2 Myr (sys.)]} & isochronal age (FGKM stars)\\
\hline
\end{tabular}
\label{tab:BPMG_ages}
\end{table*}

\citet{Barrado99} listed 7 systems in the BPMG, and quoted an age of
$20 \pm 10\,\rm{Myr}$ (based on isochronal ages for the
pre-main-sequence [pre-MS] M-type stars GJ 799AB and GJ
803). \citet{Mamajek01} used simple linear trajectories to show that
$\beta$ Pic and the M dwarfs GJ 799AB and GJ 803 may have formed in
the vicinity of the Scorpius-Centaurus (Sco-Cen) association
$\sim12\,\rm{Myr}$ ago.  Other nearby young stars showed a similar
pattern, with isochronal ages similar to their times of closest pass
to Sco-Cen, including the HD 199143 and HD 358623
(BD-17$^{\circ}$6128) binary, the V824 Ara (HD 155555) triple system,
V343 Nor, and PZ Tel.  \citet{Zuckerman01} provided a list of BPMG
members that subsumed only $\beta$ Pic, GJ 799AB and GJ 803 from the
\citet{Barrado99} list, plus the HD 199143, HD 358623, V824 Ara, V343
Nor, and PZ Tel systems mentioned by \citet{Mamajek01}, and a dozen
other systems. \citet{Zuckerman01} estimated an age of
$12^{+8}_{-4}\,\rm{Myr}$ based mostly on the position of the stars in
the Hertzsprung-Russell (H-R) diagram compared to theoretical
isochrones, and a comparison of the Li abundances to those of the TW
Hydrae Association (TWA) and Tucana-Horologium (Tuc-Hor) moving group
members. \citet{Song02} found evidence that the Li depletion boundary
(LDB) among BPMG members was around spectral type $\sim$ M4-M4.5 and
quoted an upper limit on the group age of $< 20\,\rm{Myr}$.  A more
recent study by \citet{Mentuch08} measured Li depletion amongst the
\citet{Zuckerman04} members and derived a model-dependent age of $21
\pm 9\,\rm{Myr}$ through comparison with pre-MS evolutionary
models. \citet{Yee10} reanalysed a subset of the \citet{Zuckerman01}
members and compared the ages derived from the H-R diagram and Li
depletion. They do not quote an age estimate for the BPMG, but
demonstrated that the Li depletion ages were systematically older than
the ages derived via the H-R diagram, and attributed this to the
discrepancy between the measured M dwarf radii and those predicted by
the evolutionary models. Recently, \citet{Binks14} have confirmed that
the LDB is at M4.5\,$\pm$\,0.5 ($V-K_{\rm{s}} \simeq 5.5 \pm
0.1\,\rm{mag}$ and $M_{K_{\rm{s}}} \simeq 5.6 \pm 0.4\,\rm{mag}$),
consistent with an LDB age of $21 \pm 4\,\rm{Myr}$ (statistical) and
model dependent (systematic) uncertainty of only $\pm 1\,\rm{Myr}$ (as
estimated using three sets of evolutionary tracks).

Multiple studies have estimated an expansion age of between 11 and
$12\,\rm{Myr}$ for BPMG.  \citet{Ortega02} traced back the orbits of
19 BPMG systems from \citet{Zuckerman01} using a realistic Galactic
potential, and found that the group was most concentrated within a
$\sim 24\,\rm{pc}$ wide region 11.5 Myr ago. \citet{Ortega02}
considered their age estimate a ``kinematical age'', however it is
unclear whether this age refers to the minimum mean separation between
the individual members, or the girth of the entire group as defined by
the most distant members. If it is the latter, then the age is
unlikely to be robust as it is subject to the statistical whims of
observations or the presence of interlopers. \citet*{Song03} assumed
linear trajectories and found that, excepting 3 outliers, ``{\it all
  members were confined in a smaller space about $\sim$ 12\,Mya}'',
where the region encircled in their fig. 4 is about $\sim$\,30\,pc in
diameter.  \citet{Ortega04} conducted another analysis, including the
new members from \citet{Song03}, and claim that a group of 14 of the
BPMG members reached a minimum size ($\sim 35\,\rm{pc}$ in radius)
$10.8 \pm 0.3$ Myr ago.

\citet{Torres06} claim to find evidence of expansion in the BPMG
(their fig. 7) and estimate a linear trend of:

\begin{equation} 
U = 0.053\,{\rm km\,s^{-1}\,pc^{-1}}\,(X) - 11\,{\rm km\,s^{-1}}
\end{equation}

\noindent where $U$ is in \kms\, and $X$ is in pc.  Although they do
not cite an expansion age, their slope of 0.053 \kmspc\, equates to an
expansion age of $18\,\rm{Myr}$, however no uncertainty is given.
\citet{Makarov07} tracked the times of closest encounter for 14 BPMG
members, and concluded that the mean time of nearest approach for
those pairs was $22 \pm 12$ Myr ago (the analysis excluded HIP 29964
and $\beta$ Pic itself, which was consistently giving minimum
separation encounter times with other members of $50-70$ Myr ago).
Including HIP 29964 and $\beta$ Pic in his analysis, \citet{Makarov07}
found the mean time of minimum separation to be $31 \pm 21$ Myr ago.
\citet{Makarov07} concluded that the expansion was weak ($\sim 2-3$
\kms) and that the scatter of encounter times was fairly large. The
subsequent kinematic studies by \citet{Torres06} and \citet{Makarov07}
are suggestive that the $11-12\,\rm{Myr}$ expansion age quoted by the
earlier studies may be irreproducible, or at least depend critically
upon which sample of BPMG members are included in the kinematic
analysis.

The strength of the conclusions from \citet{Ortega02}, \citet{Song03},
and \citet{Ortega04} have caused some \cite*[e.g.][]{Song12} to claim
that the expansion age of $12\,\rm{Myr}$ for BPMG is ``model
independent'' and therefore more reliable than other age
methods. However, this conclusion appears to have ignored the findings
of \citet{Torres06} and \citet{Makarov07}, which are consistent with
much slower expansion, and older expansion ages.

The lithium depletion boundary (LDB) -- the luminosity at which Li
remains unburned -- in the BPMG was recently identified by
\cite{Binks14} who derived an age of $21\pm4\,\rm{Myr}$. More
recently, \cite{Malo14b} used the Dartmouth magnetic stellar
evolutionary models (see \citealp{Feiden12,Feiden13}) to derive an
average isochronal age of between 15 and $28\,\rm{Myr}$ and an LDB age
of $26\pm3\,\rm{Myr}$ using a sample of late K and M dwarfs in the H-R
diagram. Although LDB ages are derived using pre-MS evolutionary
models, the \emph{luminosity} at which stars in a presumably coeval
group transition from those that demonstrate depleted Li to those that
do not, is remarkably insensitive to variations in the input physics
adopted in the evolutionary models (see e.g. \citealp*{Burke04};
\citealp{Jeffries05}). Given the physically simplistic nature of the
models which predict the LDB, in conjunction with the high level of
model-insensitivity, \citet{Soderblom13} describe LDB ages as
``semi-fundamental". Furthermore, they recommend that LDB ages offer
the best hope of establishing a reliable and robust age \emph{scale}
for young $(< 200\,\rm{Myr})$ stellar populations. Thus, it is clear
that the revised age of $\sim 20-25\,\rm{Myr}$ represents the new
``benchmark" age for the BPMG, against which other age-dating
techniques should be validated or tested.

Given the large (factor of $\sim$2) discrepancy between the LDB and
previous kinematic ages, in this contribution we independently
reexamine the evidence for kinematic expansion of the BPMG using the
best available astrometric data, and conclude whether a unique
kinematic age can be assigned to the group. Furthermore, we also
investigate age constraints based on the ``turn-on" to the
main-sequence as observed amongst the A-, F- and G-type members, and
compare these to theoretical model isochrones in colour-magnitude
diagrams (CMDs).

\section{Data}
\label{data}

\subsection{Kinematic data}
\label{kinematic_data}

We use the BPMG membership from \citet{Zuckerman04} as it defines a
sample of ``classic'' BPMG members with \emph{Hipparcos} astrometry
used in most studies. This sample is more restrictive than the larger
sample assembled by \citet{Torres06} and \citet{Torres08}, however
these larger samples contain very few stars with trigonometric
parallaxes, and many at larger distances (which increases the chances
of interlopers polluting the sample). Ideally, the inclusion or
exclusion of a few members should {\it not} make or break whether a
group has a detectable expansion and corresponding age.

Velocities and positions in Galactic cartesian coordinates were
calculated for members of the BPMG using the best available published
astrometry and velocities (these are shown in Table \ref{tab:BPMG}). All
trigonometric parallaxes and celestial positions were taken from the
revised \emph{Hipparcos} catalogue of \citet{vanLeeuwen07}. Proper motions
were usually adopted from \citet{vanLeeuwen07}, unless more recent
long-term proper motions from the UCAC4 catalogue \citep{Zacharias12}
(which has smaller uncertainties) were available. The majority of
radial velocities come from either the compiled catalogue of
\citet{Gontcharov06} or the recent survey of \citet{Bailey12}.

\begin{table*}
\caption[]{Positions and velocities of the BPMG stars. Galactic $XYZ$
  positions and $UVW$ velocities are defined as towards the Galactic
  centre ($X,U$), $\ell$ = 90$^{\circ}$ ($Y,V$), and the North
  Galactic pole ($Z,W$). Three-dimensional velocities were calculated using
  kinematic data from the following references (astrometry reference
  listed first, radial velocity reference listed second). If three
  references are listed, the source of the parallax is listed first,
  the source of the proper motion is listed second and the source of the
  radial velocity is listed third.  Positions, spectral types, and
  $V$-band magnitudes are given in
  \protect\citet{Zuckerman04}.}
\centering
\begin{tabular}{l l c c c c c c}
\hline
Name            & Ref.   & $X$   & $Y$   & $Z$   & $U$           & $V$           & $W$\\
...             & ...    & (pc)    & (pc)    & (pc)    & (\kms)          & (\kms)          & (\kms)\\
\hline
HR      9       & 1,2    &  4.5  &   5.9 & -38.7 & -11.0$\pm$0.7 & -15.1$\pm$1.9 & -10.2$\pm$2.9\\
HIP     10679   & 1,3,4  & -19.3 &  13.6 & -13.9 & -10.9$\pm$1.9 & -10.8$\pm$1.3 &  -5.4$\pm$0.9\\
HIP     10680   & 1,3,2  & -24.3 &  17.1 & -17.5 & -12.1$\pm$1.5 & -14.4$\pm$1.1 &  -6.8$\pm$0.8\\
HIP     11437B  & 1,3,5  & -29.4 &  19.8 & -18.5 & -13.1$\pm$1.5 & -14.6$\pm$1.2 &  -7.8$\pm$0.8\\
HIP     11437   & 1,3,5  & -29.4 &  19.8 & -18.5 & -13.5$\pm$1.4 & -14.2$\pm$1.0 &  -8.3$\pm$0.6\\
HIP     12545   & 1,3,5  & -27.5 &   7.0 & -31.0 & -10.2$\pm$0.8 & -17.7$\pm$0.8 &  -6.1$\pm$0.7\\
51      Eri     & 1,6    & -24.0 &  -8.1 & -15.0 &  -7.2$\pm$0.3 & -13.9$\pm$0.2 &  -5.7$\pm$0.1\\
GJ      3305    & 1,3,5  & -24.0 &  -8.1 & -15.0 & -13.8$\pm$0.4 & -16.2$\pm$0.3 &  -9.6$\pm$0.2\\
HIP     23309   & 1,3,7  &  -1.5 & -21.2 & -16.3 & -11.2$\pm$0.3 & -16.6$\pm$0.3 &  -9.2$\pm$0.2\\
HIP     23418   & 1,3,2  & -30.9 &  -5.7 & -10.8 & -10.1$\pm$3.7 & -14.4$\pm$3.6 &  -9.4$\pm$2.5\\
HIP     35850   & 1,2    & -20.4 & -13.9 & -11.0 & -12.5$\pm$1.1 & -16.8$\pm$0.6 &  -9.5$\pm$0.3\\
$\beta$ Pic     & 1,2    &  -3.4 & -16.4 &  -9.9 & -11.0$\pm$0.5 & -16.0$\pm$0.5 &  -9.1$\pm$0.3\\
HIP     29964   & 1,3,8  &   7.4 & -33.1 & -18.2 & -10.7$\pm$0.6 & -16.1$\pm$0.6 &  -8.2$\pm$0.4\\
HIP     76629   & 1,2    &  31.1 & -22.7 &  -1.2 &  -9.2$\pm$0.9 & -17.3$\pm$0.8 &  -9.8$\pm$0.5\\
HR      6070    & 1,2    &  39.0 &  -7.9 &  11.0 & -13.6$\pm$0.6 & -16.1$\pm$0.4 & -12.3$\pm$0.3\\
HD      155555  & 1,2    &  24.7 & -17.4 &  -8.8 &  -9.5$\pm$0.8 & -16.6$\pm$1.0 &  -8.8$\pm$0.6\\
HIP     88399   & 1,2    &  44.3 & -14.7 & -11.6 &  -8.0$\pm$0.5 & -15.6$\pm$0.5 &  -9.2$\pm$0.3\\
HR      6749    & 1,3,2  &  40.4 &  -7.5 &  -7.9 & -12.3$\pm$0.5 & -15.4$\pm$0.4 &  -6.9$\pm$0.3\\
HIP     92024   & 1,2    &  22.8 & -12.8 & -11.5 & -10.7$\pm$2.4 & -15.3$\pm$3.0 &  -9.0$\pm$1.8\\
CD-64   1208    & 1,9,7  &  22.8 & -12.8 & -11.5 & -12.2$\pm$1.8 & -16.2$\pm$2.1 &  -8.6$\pm$1.3\\
PZ      Tel     & 1,3,2  &  46.8 & -11.5 & -18.2 & -11.7$\pm$0.8 & -15.2$\pm$0.6 &  -8.4$\pm$0.4\\
HR      7329    & 1,2    &  41.3 & -12.7 & -21.3 &   2.1$\pm$2.6 & -18.9$\pm$2.8 & -14.0$\pm$1.8\\
HIP     95270   & 1,2,   &  44.4 & -13.8 & -22.9 &  -9.5$\pm$0.6 & -16.6$\pm$0.5 &  -8.6$\pm$0.3\\
GJ      799A    & 1,10,5 &   8.5 &   1.7 &  -6.3 &  -9.8$\pm$0.6 & -16.6$\pm$0.5 & -11.1$\pm$0.4\\
GJ      799B    & 1,10,5 &   8.5 &   1.7 &  -6.3 & -11.5$\pm$0.7 & -18.3$\pm$0.6 & -11.5$\pm$0.4\\
GJ      803     & 1,5    &   7.7 &   1.7 &  -5.9 &  -9.8$\pm$0.2 & -16.3$\pm$0.1 & -10.7$\pm$0.1\\
HD      199143  & 1,7    &  32.3 &  18.9 & -26.2 &  -7.5$\pm$1.4 & -13.6$\pm$1.1 & -11.2$\pm$1.3\\
BD-17 6128      & 1,11,7 &  32.3 &  18.9 & -26.2 &  -9.0$\pm$0.8 & -14.2$\pm$0.7 &  -9.4$\pm$0.7\\
HIP     112312  & 1,3,5  &  10.7 &   2.4 & -20.6 & -11.5$\pm$1.6 & -17.9$\pm$1.1 & -11.5$\pm$0.6\\
HIP     112312B & 1,3,5  &  10.7 &   2.4 & -20.6 & -11.2$\pm$1.6 & -18.1$\pm$1.1 & -10.2$\pm$0.6\\
\hline
{\bf Mean}      & ...    & {\bf 8.4} & {\bf -5.0} & {\bf -15.0} & {\bf -10.9$\pm$0.3} & {\bf -16.0$\pm$\,0.3} & {\bf -9.2$\pm$0.3}\\
\hline
\end{tabular}

\vspace{1pt}
\begin{flushleft}
References: (1)
  \protect\citet{vanLeeuwen07}, (2) \protect\citet{Gontcharov06}, (3)
  \protect\citet{Zacharias12}, (4) \protect\citet{Nordstrom04}, (5)
  \protect\citet{Bailey12}, (6) \protect\citet{Song03}, (7)
  \protect\citet{Torres06}, (8) adopted average $v_r$ of
  \protect\citet{Zuckerman01} and \protect\citet{Torres06}
  (15.7$\pm$0.7 \kms, adopting 1 \kms\, error for \citeauthor{Torres06} value), (9)
  \protect\citet{Kharchenko09}, (10) \protect\citet{Hog00}, (11)
\protect\citet{Zacharias10}.
\end{flushleft}
\label{tab:BPMG}
\end{table*}

\subsection{Photometric data}
\label{photometric_data}

All the stars in our sample are in the \emph{Hipparcos} catalogue and
have Johnson $BV$ photometry (either estimated using ground-based
observations or from \emph{Hipparcos}). However, no uncertainties are
explicitly given for the Johnson $V$-band magnitudes. Instead we adopt
$BV$ photometry from the homogenised $UBV$ catalogue of
\citet{Mermilliod91}. If homogenised $BV$ data are unavailable, or
there are no quoted uncertainties, Tycho-2 $(BV)_{_{\rm{T}}}$
photometry \citep{Hog00} was instead adopted and transformed to
Johnson $BV$ using the conversions presented in
\cite*{Mamajek02,Mamajek06_erratum}. In addition to the ``classic"
\citet{Zuckerman04} sample of members as discussed in
Section~\ref{kinematic_data}, we also include 4 stars (HIP 86598, HIP
89829, HIP 92680, and HIP 99273) which are listed as new bona fide
members from the study of \citet{Malo13}.  Table~\ref{tab:BPMG_phot}
lists the compiled $BV$ photometry in addition to spectral types and
distances for the A-, F- and G-type BPMG members.

\begin{table*}
\caption[]{Spectral types, distances and photometry for A-, F- and G-type members of the BPMG.}
\centering
\begin{tabular}{l l l c c c c c l}
\hline
Name             &   SpT   &   Ref.&  $V$               &   $B-V$             &   Ref.  &    Distance         &   $M_{V}$ & Phase\\
...           &...   &... &  (mag)             &   (mag)             &...  &    (pc)             &   (mag)  &...\\
\hline                                                                                                           
HR 6070          &   A1Va  &   1   &  $4.784\pm0.009$   &   $0.016\pm0.007$   &   9,9   &    $41.29\pm0.38$   &   $1.71\pm0.05$ & ZAMS\\
HR 7329          &   A0V   &   1   &  $5.043\pm0.005$   &   $0.020\pm0.002$   &   9,10  &    $48.22\pm0.49$   &   $1.63\pm0.05$ & ZAMS\\
$\beta$ Pic      &   A6V   &   2   &  $3.845\pm0.005$   &   $0.175\pm0.005$   &   9,9   &    $19.44\pm0.05$   &   $2.40\pm0.01$ & ZAMS\\
HIP 92024        &   A7V   &   2   &  $4.780\pm0.007$   &   $0.197\pm0.004$   &   9,9   &    $28.55\pm0.15$   &   $2.50\pm0.03$ & ZAMS\\
HR 6749$^{a}$     &   A6V   &   3   &  $5.635\pm0.020$   &   $0.222\pm0.020$   &   11,11  &    $41.84\pm1.16$   &   $2.53\pm0.14$ & ZAMS\\
HR 6750$^{a}$     &   A7V   &   3   &  $5.711\pm0.020$   &   $0.234\pm0.020$   &   11,11 &    $41.84\pm1.16$   &   $2.60\pm0.14$ & ZAMS\\
51 Eri           &   F0IV  &   4   &  $5.215\pm0.009$   &   $0.283\pm0.006$   &   9,9   &    $29.43\pm0.29$   &   $2.87\pm0.05$ & ZAMS\\
HR 9             &   F3V   &   5   &  $6.172\pm0.004$   &   $0.386\pm0.005$   &   9,9   &    $39.39\pm0.59$   &   $3.20\pm0.08$ & Pre-MS*\\
HIP 88399        &   F4.5V &   5   &  $7.007\pm0.010$   &   $0.458\pm0.015$   &   11,10 &    $48.15\pm1.30$   &   $3.59\pm0.14$ & Pre-MS\\
HIP 99273$^{b}$   &   F5V   &   6   &  $7.181\pm0.010$   &   $0.480\pm0.015$   &   11,10 &    $52.22\pm1.23$   &   $3.59\pm0.12$ & Pre-MS\\
HIP 95270        &   F6V   &   5   &  $7.037\pm0.010$   &   $0.480\pm0.004$   &   11,10 &    $51.81\pm1.75$   &   $3.47\pm0.17$ & Pre-MS\\
HIP 10680$^{c}$   &   F7V   &   5   &  $7.031\pm0.011$   &   $0.518\pm0.021$   &   11,10 &    $37.62\pm2.73$   &   $4.15\pm0.36$ & ?\\
HD 199143A$^{a}$  &   F7V   &   5   &  $7.323\pm0.012$   &   $0.532\pm0.016$   &   11,11 &    $45.66\pm1.61$   &   $4.03\pm0.18$ & Pre-MS\\
HIP 25486        &   F7V   &   5   &  $6.296\pm0.010$   &   $0.551\pm0.013$   &   11,11 &    $27.04\pm0.35$   &   $4.14\pm0.07$ & Pre-MS\\
HIP 86598$^{b}$   &   F9V   &   7   &  $8.357\pm0.017$   &   $0.558\pm0.019$   &   11,11 &    $72.46\pm4.57$   &   $4.06\pm0.32$ & Pre-MS\\
HIP 10679$^{c}$   &   G3V   &   5   &  $7.755\pm0.013$   &   $0.622\pm0.008$   &   11,10 &    $37.62\pm2.73$   &   $4.88\pm0.36$ & ?\\
HIP 89829$^{b}$   &   G3IV  &   5   &  $8.797\pm0.019$   &   $0.639\pm0.031$   &   11,11 &    $72.57\pm5.37$   &   $4.49\pm0.37$ & Pre-MS\\
HD 155555A$^{a,d}$ &  G5V    &  5   &  $7.222\pm0.100$   &   $0.743\pm0.100$   &   12,10  &   $31.45\pm0.49$   &    $4.73\pm0.13$ & Pre-MS\\
HIP 92680$^{b}$   &   G9IV  &   8   &  $8.461\pm0.014$   &   $0.780\pm0.008$   &   9,9   &    $51.49\pm2.60$   &   $4.90\pm0.25$ & Pre-MS\\
\hline
\end{tabular}

\vspace{1pt}
\begin{flushleft}
References: 
(1) \protect\cite{Gray87}, 
(2) \protect\cite{Gray06}, 
(3) \protect\cite{Corbally84},
(4) \protect\cite{Gray89},
(5) \protect\cite{Pecaut13}, 
(6) \protect\cite{Houk82},
(7) \protect\cite{McCarthy12}, 
(8) \protect\cite{Torres06}, 
(9) \citet{Mermilliod91},
(10) \citet{Perryman97},
(11) Tycho $(BV)_{\rm{T}}$ photometry transformed to Johnson $BV$ following \citet{Mamajek06},
(12) \citet{Strassmeier00}. 
Distances are based on parallax measurements from the revised
\emph{Hipparcos} reduction of \protect\citet{vanLeeuwen07}.\\

Evolutionary phases are based on CMD positions: ZAMS = zero-age main-sequence,
Pre-MS* = pre-main-sequence phase between penultimate and final (ZAMS) luminosity minima
(see Section~\ref{colour-magnitude_diagram}), Pre-MS = pre-main-sequence.\\
Notes:\\
$^{a}$ Individual binary components shown (see Section~\ref{colour-magnitude_diagram} for a description of deconstructing the combined photometric measurements).\\
$^{b}$ Not listed as members in \protect\citet{Zuckerman04}, but identified as bona fide members in \protect\citet{Malo13}.\\
$^{c}$ Due to a large uncertainty in the parallax measurement we instead adopt a kinematic distance (see Section~\ref{colour-magnitude_diagram}).\\
$^{d}$ The large uncertainty on both the magnitude and colour is due to the high level of variability associated with the object. We therefore adopt the mean $V$-band magnitude as given by \protect\cite{Strassmeier00} and deconstruct the combined photometry as detailed in Section~\ref{colour-magnitude_diagram}.
\end{flushleft}
\label{tab:BPMG_phot}
\end{table*}

%%% ANALYSIS SECTIONS %%%

\section{Kinematic analysis}
\label{analysis}

The following sections are laid out, in order to test in as simple a
way as possible, whether or not it is possible to assign a unique
kinematic age to the BPMG.
In Section~\ref{revised_kinematic_parameters_bpmg}, we calculate
revised basic kinematic parameters for the BPMG (mean velocity and
velocity dispersion).
In Section~\ref{velocity_trends_consistent_with_expansion}, we
calculate the velocity gradients and statistically test whether they
are consistent with expansion ages of $11-12\,\rm{Myr}$.
In Section~\ref{simple_linear_trajectories}, we reattempt the analysis
of \citet{Song03} using contemporary kinematic data, following the
linear trajectories of the BPMG stars, in search for a signature of an
expansion age.
In Section~\ref{epicycle}, we model the orbits using
epicycle approximation, and examine the spatial distribution in the
past in search of a unique expansion age. 

Astronomical samples often have interlopers, so calculating
statistical means, standard errors of means, and standard deviations
can often be biased by such interlopers \citep[e.g.][]{Gott01}.
Throughout the kinematic analysis, when we mention a ``mean'', we are
actually using the average of (1) the median, (2) the Chauvenet
clipped mean \citep{Bevington92}, and (3) the probit mean
\citep{Lutz80}, i.e.  a ``mean'' that is remarkably immune to the
effects of outliers. When we mention a ``standard error'', we are
actually using the average of (1) the error of the true median
\citep{Gott01}, and (2) the standard error of the Chauvenet clipped
mean \citep{Bevington92}.  When we mention ``dispersion'', we are
quoting the average of (1) the 68 per cent confidence intervals, and (2)
the probit standard deviation \citep{Lutz80}. In the limit of a
Gaussian distribution lacking outliers, these quantities are
asymptotically identical to their normal counterparts: the mean,
standard error of the mean, and standard deviation. Our choice of
estimators allows one to calculate the statistical moments in a way
that is not sensitive to the inclusion or exclusion of some
``problem'' stars.  Such stars may have been arbitrarily included or
excluded in previous studies.

%%% REVISED KINEMATIC PARAMETERS

\subsection{Revised kinematic parameters for the BPMG}
\label{revised_kinematic_parameters_bpmg}

Based on the three-dimensional velocities for the 30 BPMG members in Table
2, we estimate a new mean velocity for the BPMG of:

\begin{equation}
U,\,V,\,W\,=\,-10.9\,\pm\,0.3, -16.0\,\pm\,0.3, -9.2\,\pm\,0.3\,{\rm km\,s^{-1}}
\end{equation}

\noindent which is in good agreement with the recent estimate of
\citet{Malo14a}. Here the Galactic $U, V, W$ velocities are defined in
the classic sense, and $U$ is explicitly positive toward the Galactic
centre. This space motion corresponds to an approximate convergent
point solution of:

\begin{equation}
\alpha_{\mathrm{CP}}\, = 88^{\circ}.0\,\pm\,0^{\circ}.9
\end{equation}
\begin{equation}
\delta_{\mathrm{CP}}\, = -30.6^{\circ}\,\pm\,0^{\circ}.8
\end{equation}
\begin{equation}
S_{\mathrm{tot}}\, = 21.4\,\pm\,0.3~{\rm km\,s^{-1}}
\end{equation}

The {\it observed} 1$\sigma$ dispersions are $\sigma_U$, $\sigma_V$,
$\sigma_W$ = 1.8, 1.6, and 1.9 \kms, respectively. However to estimate
the {\it intrinsic} one-dimensional velocity dispersion, we need to take into
account the mean uncertainties in the individual $U, V, W$ velocities
(0.9, 0.9, 0.6 \kms, respectively). Hence, subtracting off the mean
errors in quadrature, we estimate the {\it intrinsic one-dimensional velocity
  dispersions} for the BPMG $U, V, W$ velocities to be:

\begin{equation}
\sigma_U^{\mathrm{int}}, \sigma_V^{\mathrm{int}}, \sigma_W^{\mathrm{int}}\, = 1.5,~1.4,~1.8~{\rm km\,s^{-1}}
\end{equation}

\noindent respectively, i.e. $\sim$1.5 \kms\, in all directions.  This
is similar to that calculated for the subgroups in the Sco-Cen complex
(e.g. \citealp*{deBruijne01}; \citealp{Mamajek03}).  The
one-dimensional velocity dispersions are somewhat similar to those
calculated by \citet{Malo13} 
($\sigma_U^{\rm{int}}$, $\sigma_V^{\rm{int}}$, $\sigma_W^{\rm{int}}$\,
= 2.06, 1.30, and 1.54\,$\rm{km\,s^{-1}}$) and \citet{Malo14a}
($\sigma_U^{\rm{int}}$, $\sigma_V^{\rm{int}}$, $\sigma_W^{\rm{int}}$\,
= 2.06, 1.32, and 1.35\,$\rm{km\,s^{-1}}$), which used larger samples.
However, it is unclear from their text whether the one-dimensional
dispersions in \citet{Malo13} are simply $1 \sigma$ scatters in the
observed velocities (and hence, not intrinsic velocity dispersions) or
whether it includes observational uncertainties.  The velocity
dispersions that we have listed do not take into account any
position-dependence of the velocities (which would be a manifestation
of expansion, or Galactic differential rotation). This {\it intrinsic}
one-dimensional velocity dispersion should be kept in mind when
assigning membership to the BPMG for newly identified young field
objects, especially those with heteroskedastic velocity uncertainties.

\subsection{Are the velocity trends consistent with expansion?}
\label{velocity_trends_consistent_with_expansion}

\begin{figure}
\centering
\includegraphics[width=\columnwidth]{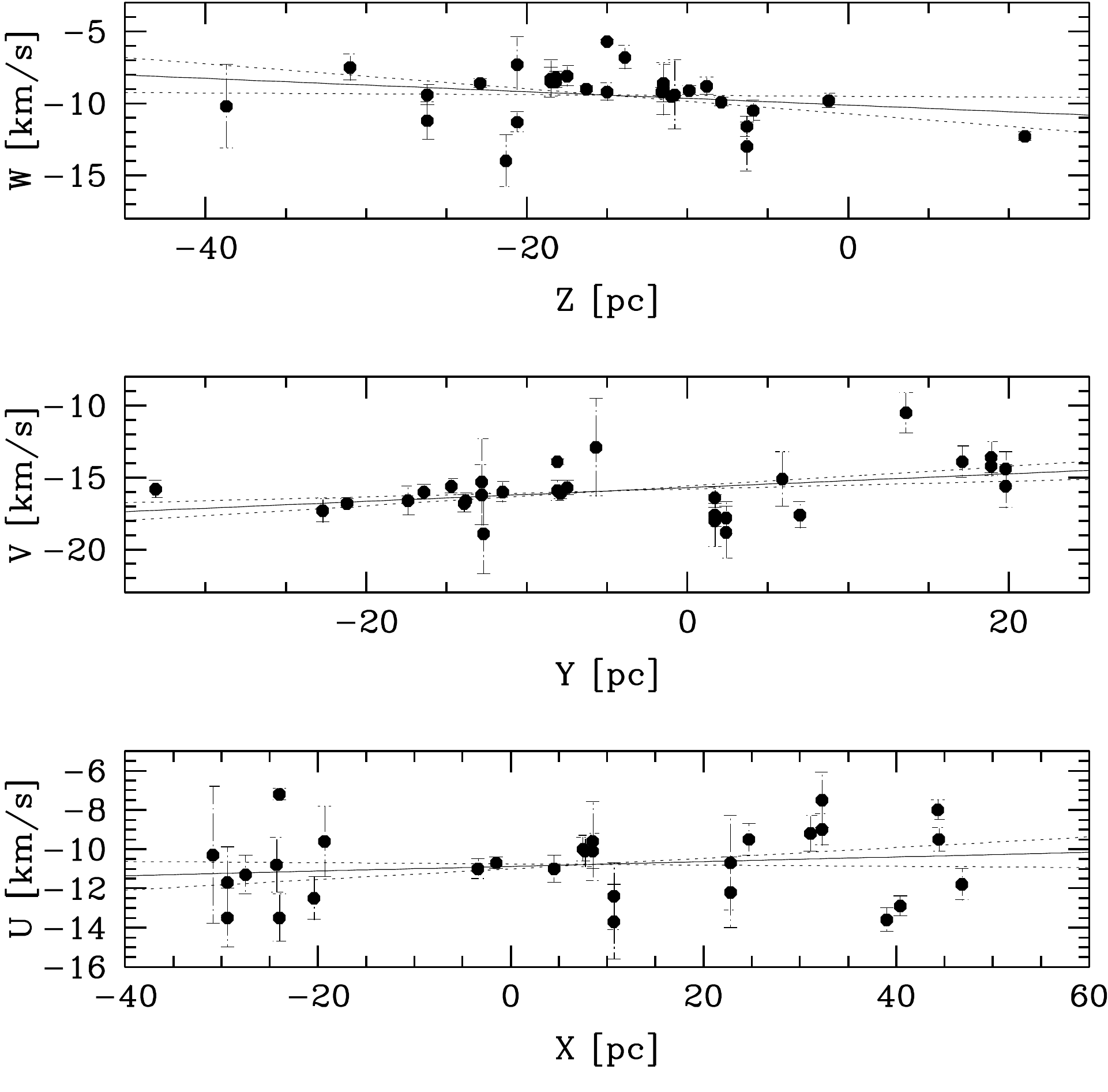}
\caption[]{These three panels show position vs. velocity along the 3
  Galactic cartesian axes for the BPMG members from
  \citet{Zuckerman04} with measured trigonometric parallaxes:
    Bottom panel: $X$ vs. $U$ (= d$X$/d$t$). Middle panel:  $Y$ vs. $V$ (=
  d$Y$/d$t$). Top panel: $Z$ vs. W (= d$Z$/d$t$).}
\label{fig:XYZUVW}
\end{figure}

Using the positions and velocities for the BPMG members compiled in
Table \ref{tab:BPMG}, we plot the position coordinates $XYZ$ vs. the
velocities along those axes $UVW$ in Fig. \ref{fig:XYZUVW}. We use
bootstrap sampling of unweighted ordinary least squares fits to
calculate the slopes and slope uncertainties for the $UVW$ velocity
vs. position trends.  For the 30 BPMG stars in Table \ref{tab:BPMG},
we estimate:

\begin{eqnarray}
\kappa_X = \frac{dU}{dX} = +0.039 \pm\, 0.024\, {\rm km\,s^{-1}\,pc^{-1}}\\
\kappa_Y = \frac{dV}{dY} = +0.052 \pm\, 0.019\, {\rm km\,s^{-1}\,pc^{-1}}\\
\kappa_Z = \frac{dW}{dZ} = -0.031 \pm\, 0.044\, {\rm km\,s^{-1}\,pc^{-1}}
\end{eqnarray}

% k_X = 0.039+-0.024  => 25.6(15.9-66.7)  Myr (1sig) = 26+41-10 Myr
%                     => 25.6(11.5-111.1) Myr (2sig) = 26+86-14 Myr
% k_Y = 0.052+-0.019  => 19.2(14.1-30.3)  Myr (1sig) = 19+11-5 Myr
%                     => 19.2(11.1-71.4)  Myr (2sig) = 19+52-8 Myr
% wtav = 0.047+-0.015 => 21.3(16.1-31.3)  Myr (1sig) = 21+10-5 Myr
%                     => 21.3(13.0-58.8)  Myr (2sig) = 21+38-8 Myr

\noindent Both $\kappa_X$ and $\kappa_Z$ are within 1.7$\sigma$ of zero,
and $\kappa_Y$ is only marginally positive ($2.7 \sigma$).  The
$\kappa$ slopes have Pearson $r$ correlation coefficients of 0.35,
0.44, -0.17, and probabilities of zero correlation of 5.6, 1.6,
and 35.8 per cent, respectively.

Expansion timescales can be calculated using $\kappa$\, following the
expression $\kappa$ = $\gamma^{-1}$$\tau^{-1}$, where $\gamma$ is the
constant 1.022712165 s\,pc\,km$^{-1}$\,Myr$^{-1}$. It is unclear
whether the $\kappa_Z$ rate is actually useful given that the expected
age of the group is not insignificant compared to the vertical
oscillation period in the local Galactic disc ($\sim 80\,\rm{Myr}$).
Given the complication in interpreting the $\kappa_Z$ value, we omit
it from further consideration. The expansion rates $\kappa_X$ and
$\kappa_Y$ are independently consistent with expansion ages in $X$ and
$Y$ of $26\,\rm{Myr}$ ($^{+41}_{-10}$ 1$\sigma$; $^{+86}_{-14}$
2$\sigma$) and $19\,\rm{Myr}$ ($^{+11}_{-5}$ 1$\sigma$; $^{+52}_{-8}$
2$\sigma$), respectively.  The weighted mean expansion rate is then
$\kappa$ = 0.047\,$\pm$\,0.015 km\,s$^{-1}$\,pc$^{-1}$, which
translates to an expansion age of $21\,\rm{Myr}$ ($^{+10}_{-5}$
1$\sigma$; $^{+38}_{-8}$ 2$\sigma$).

The new estimate of $\kappa_X$ is within the uncertainty of the slope
(0.053 \kmspc) measured by \citet[][]{Torres06}. The newly derived
$\kappa_X$ depends on improved astrometry \citep{vanLeeuwen07} and a
larger sample of BPMG stars with trigonometric parallaxes than that
used by \citet{Torres06}, so it should be an improved estimate of this
velocity trend.  If one reverses the problem and {\it assumes} that the
BPMG manifested a linear expansion consistent with an expansion age of
$12\,\rm{Myr}$, one would expect a slope of $\kappa =
0.081$\,\kmspc. This differs from our estimate at the 2.3$\sigma$
level. Given the slope uncertainties, {\it a linear expansion of the
  BPMG with timescale 12\,Myr should have been statistically detected
  at the 5.4$\sigma$ and 3.9$\sigma$ level in $X$ and $Y$,
  respectively, however it clearly was not}. Hence, the velocity
trends are inconsistent with expansion on a $12\,\rm{Myr}$ timescale.
This methodology hints that the group may have been somewhat more
compact in the past, however the 95 per cent confidence range of expansion
age ($13-59\,\rm{Myr}$), and the fact that the expansion in $X$ and $Y$
together is only $\sim 3.1 \sigma$ away from zero, suggests that this
sort of methodology will not provide an accurate or useful kinematic
age with the available astrometry.

%%% LINEAR TRAJECTORIES

\subsubsection{Simple linear trajectories}
\label{simple_linear_trajectories}

\begin{figure}
\centering
\includegraphics[width=\columnwidth]{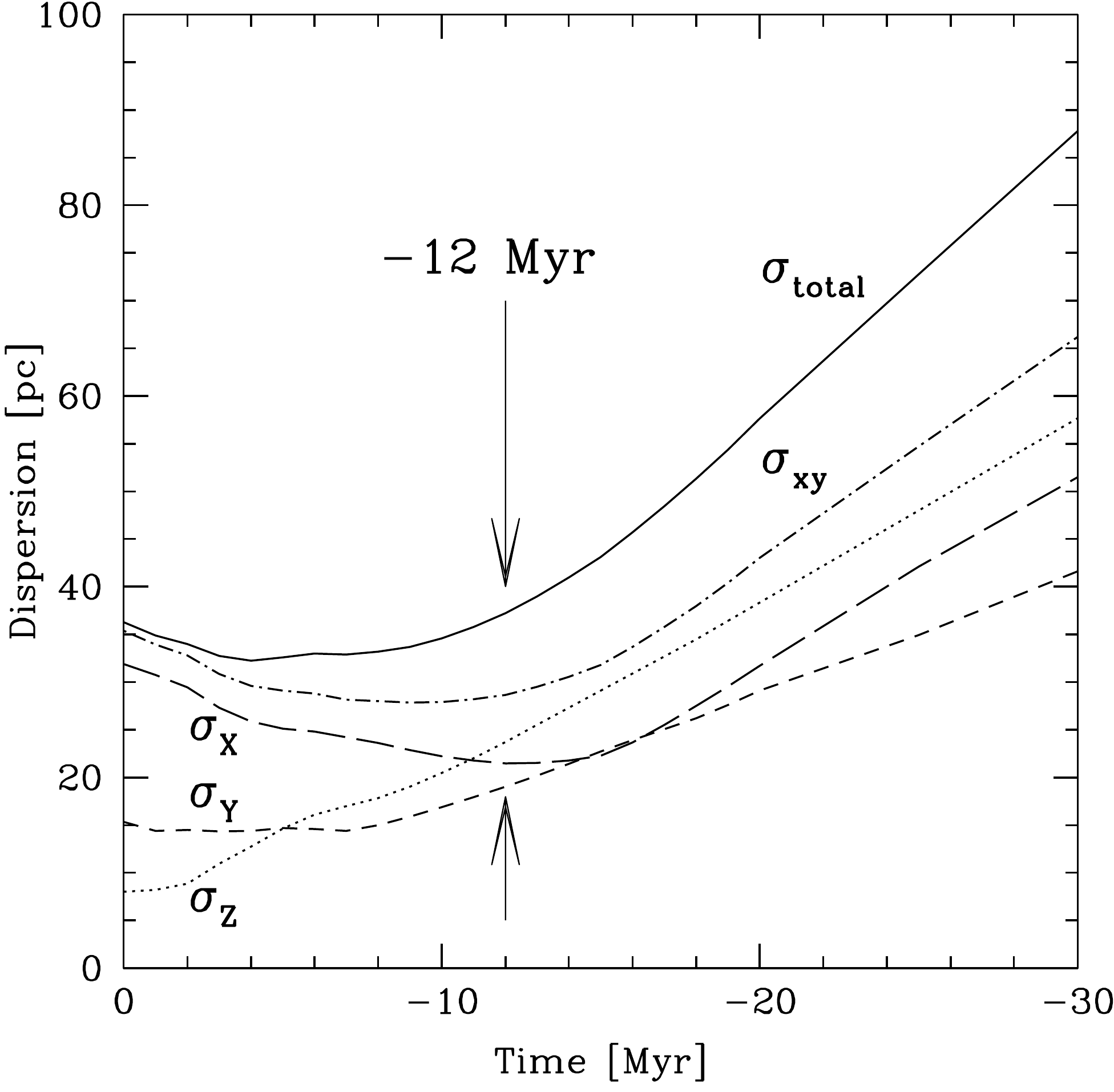}
\caption[]{1$\sigma$ dispersions in $X, Y, Z$ coordinates ($\sigma_X$,
  $\sigma_Y$, $\sigma_Z$) as a function of time in the past, assuming
  linear trajectories. The quadrature sums of the $X$- and $Y$-
  dispersions ($\sigma_{XY}$) and $X$-, $Y$- and $Z$- dispersions
  ($\sigma_{\mathrm{total}}$) are also plotted. Linear trajectories in
  $Z$ are obviously the poorest approximation (contrast with
  dispersion measured for epicyclic orbit in Fig.~\ref{fig:disp}).
  The $\sigma_{XY}$ dispersion may be the most useful overall
  metric of the group's size using the linear trajectory technique.}
\label{fig:XYZ_linear}
\end{figure}

Following \citet{Song03}, we calculate linear trajectories for the
revised positions and velocities for BPMG members in Table 2 using:

\begin{equation}
X(t) = X_{\circ} + \gamma U t
\end{equation}
\begin{equation}
Y(t) = Y_{\circ} + \gamma V t
\end{equation}
\begin{equation}
Z(t) = Z_{\circ} + \gamma W t
\end{equation}

\noindent where subscript $_{\circ}$ denotes the present positions,
$t$ is time in Myr (negative for the past), $XYZUVW$ carry the usual
definitions, and $\gamma$ is the constant 1.022712165
s\,pc\,km$^{-1}$\,Myr$^{-1}$. Fig.~\ref{fig:XYZ_linear} shows the
1$\sigma$ dispersions in $X, Y, Z$ from the present to 30 Myr ago. The
dispersion in $Z$ steadily increases as one goes further in the past.
The dispersion in $Y$ has been fairly steady over the past
$10\,\rm{Myr}$, but steadily increases as one goes further back. The
dispersion in $X$ decreases slightly as one goes back, bottoming out
at around 12 Myr ago, however the $X$ dispersion varied little ($< 5$
per cent) between 9 and 15 Myr ago.  The minima in the dispersion in
$X$, $Y$, $Z$, $X$ \& $Y$, and $XYZ$, occurred at times of 12, 3, 0,
9, and 6 Myr ago, respectively.  The minimum for the $Z$ dispersion at
present is unsurprising given how poor the approximation of a linear
trajectory is in the $Z$ direction (i.e. epicyclic period in $Z$ is
shortest among the three directions).  Arguably the most useful
dispersion to look at for the linear trajectory approximation is the
quadruture sum of $X$ and $Y$, however this shows a very broad minimum
which changed at the $<5$ per cent level between $\sim 5$ and $\sim
12$ Myr ago. We conclude that using the linear trajectory
approximation and defining the size of the group in terms of their
1$\sigma$ dispersions does not lead to an unambiguous kinematic age
for the BPMG using the currently available astrometry.

%%% EPICYCLE CODE 

\subsubsection{Epicycle orbit approximation \label{epicycle}}

\begin{figure}
\centering
\includegraphics[width=\columnwidth]{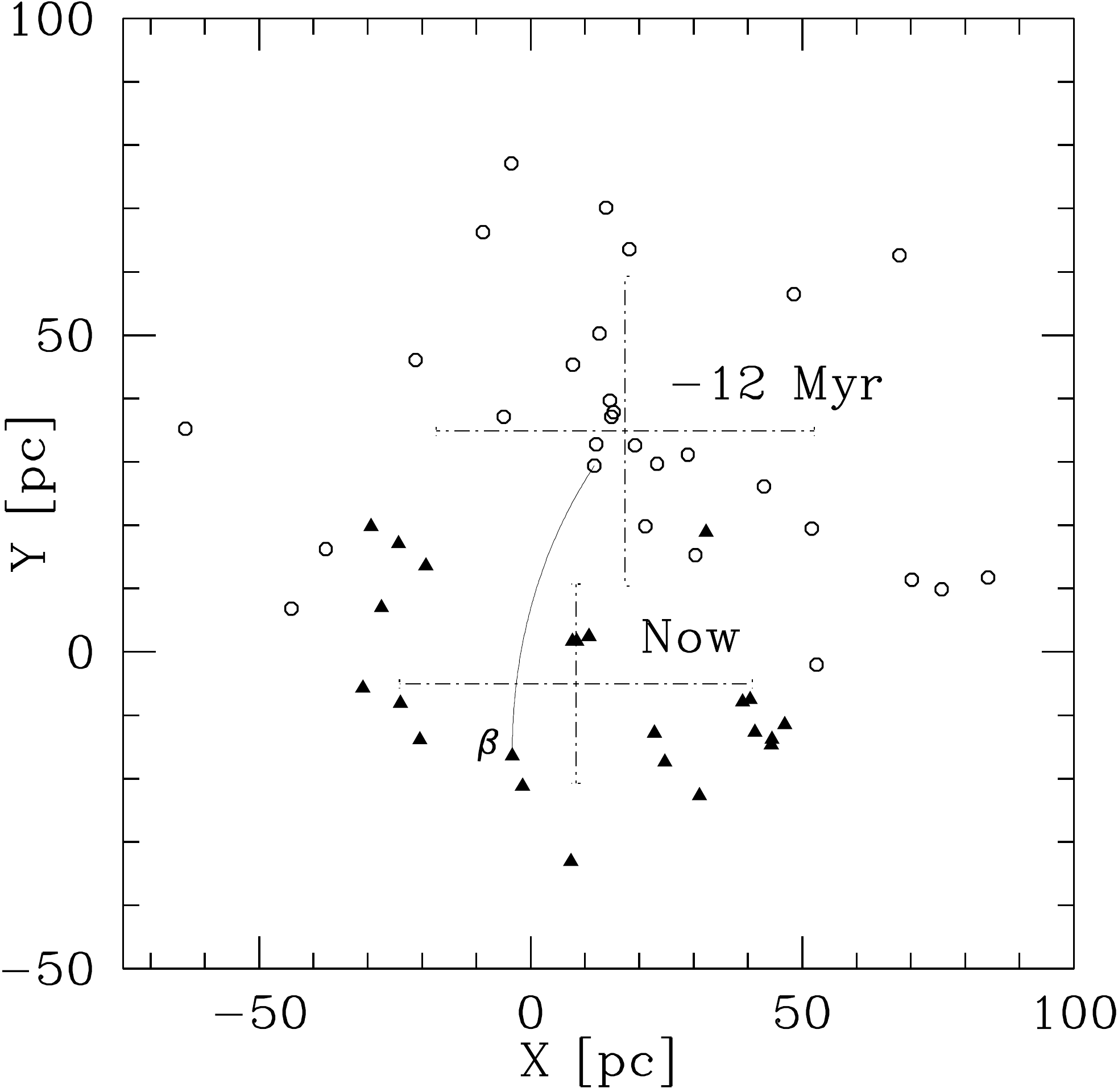}
\caption[]{Distribution of BPMG members in the $XY$ plane now ({\it
    filled triangles}) and 12 Myr ago ({\it open circles}) using
  epicycle orbit approximation.  The dispersion in the $X$ and $Y$
  directions are plotted now and 12 Myr ago. The trajectory for the
  star $\beta$ Pic itself is plotted as a solid arc, and labelled with
  a ``$\beta$''. The reference frame has its origin at the Sun's
  current position, but is co-moving with the LSR of
  \citet{Schonrich10}.}
\label{fig:past}
\end{figure}

For the epicyclic approximation, we use the orbit equations from
\citet{Fuchs06}, and adopt Oort A and B constants from
\citet{Feast97}, the local disc density from \citet{vanLeeuwen07}, the
Local Standard of Rest velocity from \citet*{Schonrich10}, and a solar
$Z$ distance above the Galactic plane of $20\,\rm{pc}$. In
Fig. \ref{fig:past}, we show the $XY$ positions for the BPMG members
from \citet{Zuckerman04} now and at 12 Myr ago. The current
distribution of BPMG members is very irregular, with no real
concentration.  It appears that 12 Myr ago, there were a few more
stars near the centroid of the group, but surprisingly the dispersion
in the positions was somewhat {\it larger} than measured for the
present. Note that the dispersions are measured using the average of
the 68 per cent confidence intervals and probit estimate of the
standard deviation, hence they are extremely resilient to the effects
of outliers.

\begin{figure}
\centering
\includegraphics[width=\columnwidth]{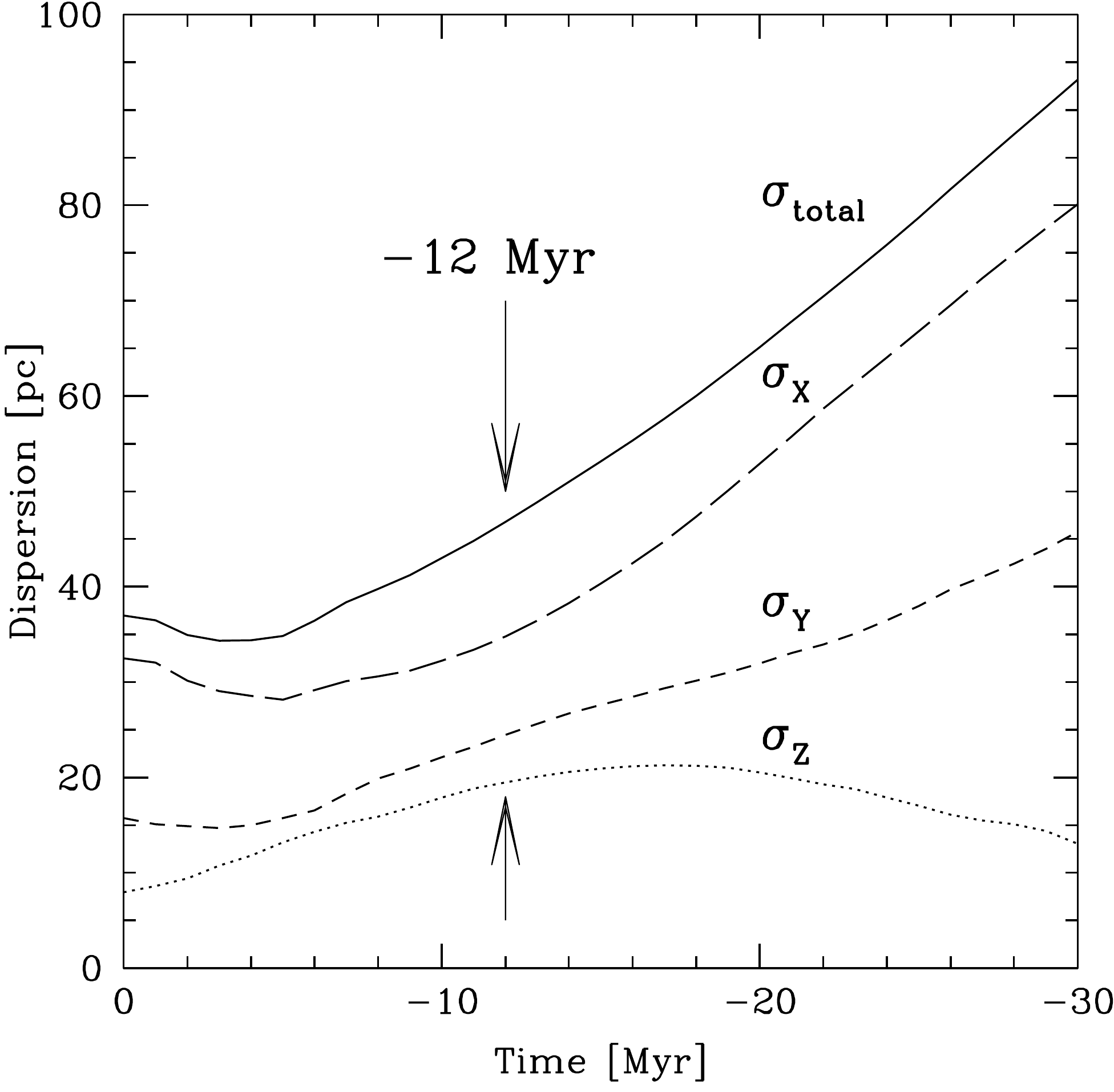}
\caption[]{Current and past dispersions in the Galactic $X$, $Y$, and
  $Z$ positions for the 30 BPMG stars from \citet{Zuckerman04}, with
  the same axes and scale as Fig. \ref{fig:XYZ_linear}. Orbits were
  traced back using an epicycle code, and the dispersions are an
  average of the 68 per cent confidence intervals and probit estimate
  of $\sigma$ (i.e. an estimate of the standard deviation immune to
  outliers). No unambiguous minimum in the dispersions is visible at
  the oft-cited age of $12\,\rm{Myr}$.}
\label{fig:disp}
\end{figure}

In Fig.~\ref{fig:disp}, we plot the dispersions in the $X, Y, Z$
directions ($\sigma_X$, $\sigma_Y$, $\sigma_Z$) as a function of time
in the past for the BPMG sample. We also plot the square root of the
sum of these dispersions added in quadrature ($\sigma_{\rm{total}}$)
as a simple metric of the overall positional dispersion. The
dispersion in the $X$ direction was marginally smaller $\sim 5$ Myr
ago, but grows monotonically further back in time (as does the
dispersion in $Y$).  The local minimum at 5 Myr ago cannot be taken
seriously as an expansion age as the velocity trends calculated in
Section~\ref{velocity_trends_consistent_with_expansion} are all
clearly inconsistent with such a rapid expansion ($5\,\rm{Myr}$
$\rightarrow$ $\kappa$ = 0.20 \kmspc). The dispersion in $Z$ is only
$\pm 8\,\rm{pc}$ now, but was $\pm 20\,\rm{pc}$ $12$ Myr ago.
Surprisingly, we find that the size of the group, as measured by a
statistical estimate of the dispersion, was somewhat {\it larger} 12
Myr ago, at odds with the claims of \citet{Ortega02} and
\citet{Song03}. As can be seen in Fig.  \ref{fig:disp}, the positional
dispersion along the 3 axes for the BPMG stars 12 Myr ago was similar
to, or somewhat larger than, the currently measured dispersion. Again,
the evidence is lacking for a more compact configuration for the BPMG
$11-12$ Myr ago, or for assigning any unambiguous expansion age.

%%% FINDING NEMO

\subsubsection{NEMO orbit integration \label{NEMO}}

\begin{figure}
\centering
\includegraphics[width=\columnwidth]{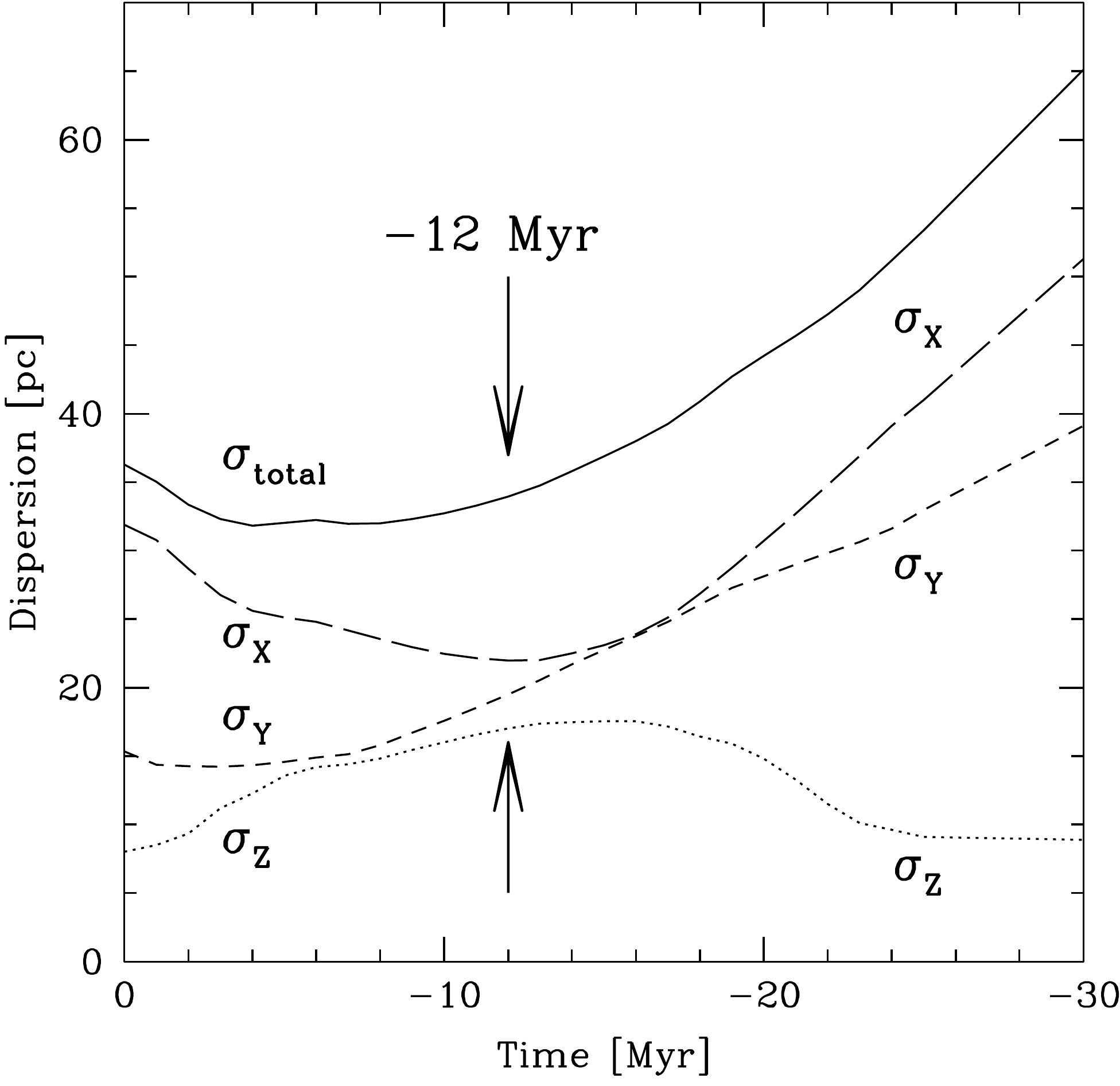}
\caption[]{Same as Fig.~\ref{fig:disp} but with the orbits integrated
  using the NEMO dynamics code. A minimum in the dispersions is not
  visible at the classic age of 12\,Myr.}
\label{fig:disp_NEMO}
\end{figure}

To determine whether a lack of unambiguous expansion age is the result
of our choice of epicycle code, we also study the past orbits of the
BPMG members using the NEMO stellar dynamics code \citep{Teuben95},
for which we adopt the Galactic potential (model 2) of
\citet{Dehnen98}.  Fig.~\ref{fig:disp_NEMO} shows the dispersions in
the $X, Y, Z$ directions as a function of time (also listed in
Table~\ref{tab:NEMO}).  While there is a minimum in $X$ positions at
$\sim 12$ Myr ago, the BPMG was larger in both $Y$ and $Z$ during that
epoch. Using the most realistic orbit approximation of the three
tried, we see that the overall dispersion in $X, Y, $ and $Z$ has a
broad minimum at between $\sim 3$ and $\sim 12$ Myr ago, but there
does not appear to be unambiguous evidence of a well-defined minimum
which could provide a useful kinematic age.

\begin{table}
\caption[]{Dispersion in positions of BPMG members now and in the past
using the NEMO stellar dynamics code.}
\centering
\begin{tabular}{c c c c c}
\hline
Time & $\sigma_X$ & $\sigma_Y$ & $\sigma_Z$\\
(Myr) & (pc) & (pc) & (pc)\\
\hline
0  & 31.9  & 15.4 & ~8.0\\
-5  & 25.1  & 14.6 & 13.6\\
-10 & 22.5  & 17.6 & 16.0\\
-12 & 22.0  & 19.5 & 17.0\\
-15 & 23.1  & 22.7 & 17.5\\
-20 & 30.7  & 28.1 & 14.8\\
-25 & 41.0  & 33.0 & ~9.1\\
-30 & 51.3  & 39.1 & ~8.9\\
\hline
\end{tabular}
\label{tab:NEMO}
\end{table}

\subsection{Discussion on kinematic ages}
\label{discussion_kinematic}

An age based on the expansion of the BPMG would only be useful if its
uncertainty were much less than the actual age.  Based on our
analysis, it is difficult to see how an expansion age with small
uncertainties can be inferred from the currently available kinematic
data. While the velocity trends in Galactic $X$ and $Y$ directions are
consistent with an expansion age of 21$^{+10}_{-5}$ Myr (1$\sigma$),
when the past orbits of the BPMG were examined using epicyclic orbit
approximation, and orbit integration using a realistic Galactic
potential, a well-defined expansion age is not apparent.  We conclude
that using the current astrometric data, we can not assign a unique,
precise expansion age.

To infer a systematic expansion age of $11-12\,\rm{Myr}$ for the BPMG,
one would probably have to choose and omit particular objects
\citep[as did][]{Song03}, however this would either negate the
reliability of determining a global expansion rate for the whole
group, or force one to significantly cut down on the number of BPMG
bona fide members (which there seems to be little astrophysical
motivation for doing). Indeed, the trend in the literature has been to
add more and more members to the BPMG, despite the fact that it
appears that the group was not significantly much more compact in the
past.

%The positions of the BPMG stars in Fig. \ref{fig:disp} \citep[and seen
%  in other published plots, e.g. ][]{Torres08} are suggestive of
%considerable substructure -- indicating that the BPMG may represent
%the remnants of an ensemble of small star-formation events with
%similar ages and velocities (and none apparently large enough to
%produce an appreciable population of OB stars).  The TWA manifests
%similar filamentary structure, hints of slow expansion, and contains
%no obvious remnant cluster \citep{Mamajek05}, all similar to the BPMG.

\section{Isochronal age from the main-sequence ``turn-on''}
\label{alternative_age_derivations_for_bpmg}

Having demonstrated that attributing an unambiguous kinematic age of
$\sim 12\,\rm{Myr}$ to the BPMG is not supported by current kinematic
evidence, we now move on to discussing age constraints from the
main-sequence ``turn-on'' and comparing the positions of stars in the
CMD to theoretical model isochrones.

For this comparison we will concentrate solely on the
intermediate-mass A-, F- and G-type BPMG members, which are likely to
either be on the zero-age main-sequence (ZAMS) or in the pre-MS
phase. We focus on the intermediate-mass stars as evolutionary models
predict multiple maxima/minima in the luminosities of the stars as
they approach the ZAMS (e.g. \citealp{Iben65}), resulting in
inflection points in the model isochrones which should be discernible
with precise data.

\subsection{Colour-magnitude diagram and pre-MS isochronal age}
\label{colour-magnitude_diagram}

\begin{figure*}
\centering
\includegraphics[width=\textwidth]{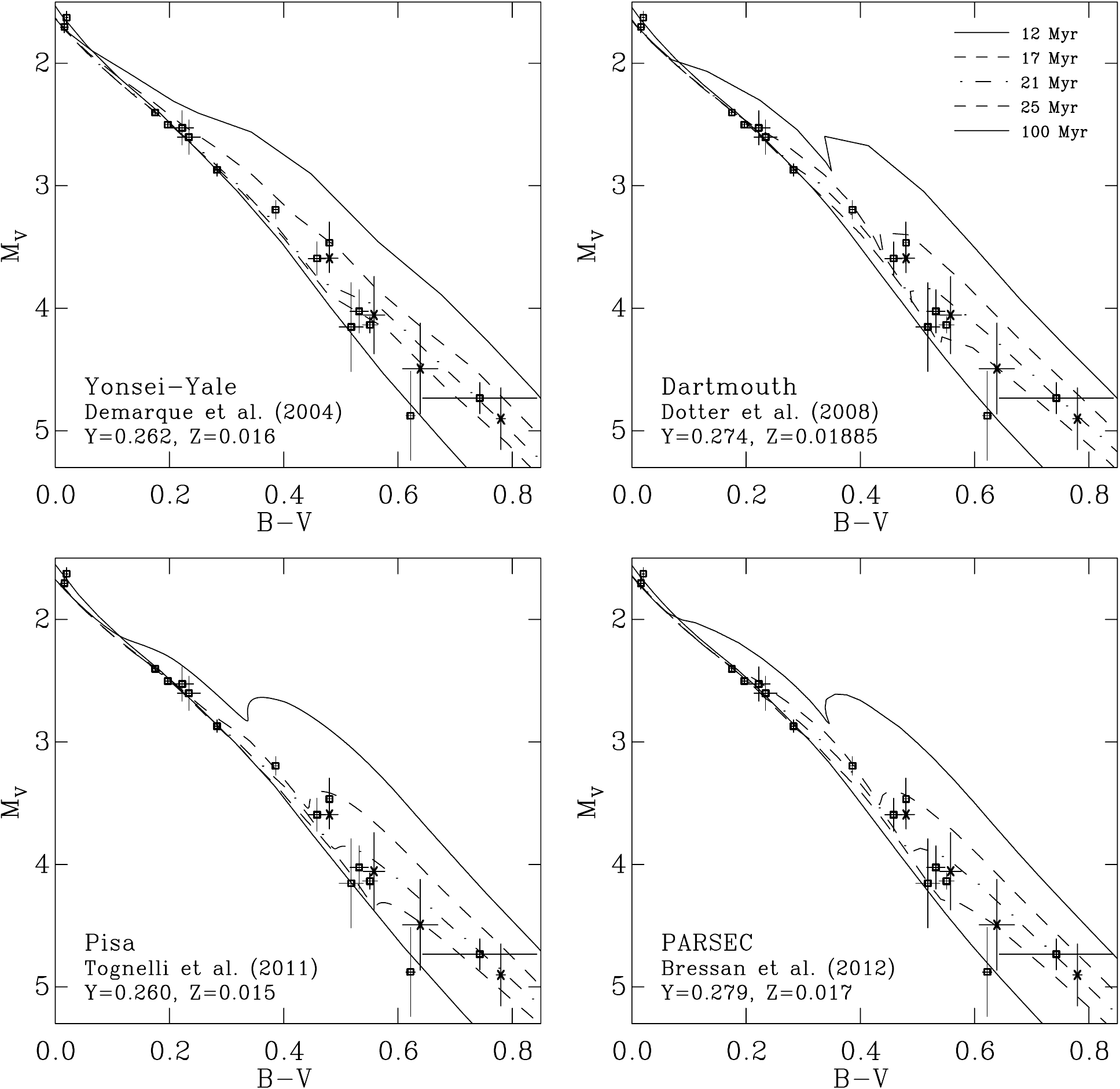}
\caption[]{$M_{V}, B-V$ CMDs of the A-, F- and G-type BPMG members
  compared against the Yonsei-Yale (Y$^{2}$; \citealp{Demarque04})
  (\emph{top left}), Dartmouth \citep{Dotter08} (\emph{top right}),
  Pisa \citep{Tognelli11} (\emph{bottom left}) and PARSEC
  \citep{Bressan12} (\emph{bottom right}) model isochrones. In all
  panels the upper continuous line represents the position of the
  single-star sequence for the oft-quoted age of $12\,\rm{Myr}$. Below
  this, the dot-dash and bounding dashed isochrones represent the
  position based on the LDB age of $21\pm4\,\rm{Myr}$ according to
  \protect\cite{Binks14}. Finally, the lower continuous line denotes
  the position for an age of $100\,\rm{Myr}$.  The squares represent
  the ``classic'' sample of members as defined by
  \protect\cite{Zuckerman04} whereas the crosses denote additional
  members from \protect\cite{Malo13}.}
\label{fig:beta_pic}
\end{figure*}

As discussed in Section~\ref{photometric_data}, we have compiled
Johnson $BV$ photometry for all the A-, F- and G-type BPMG members and
can therefore compare their positions in CMDs to theoretical model
isochrones. The reason we compare the models and the data in the
$M_{V}, B-V$ CMD is that reliable $BV$ photometry is available for all
of the stars, whereas reliable magnitudes in other common optical-IR
bands [e.g. $(RI)_{\rm{c}}JHK_{\rm{s}}$] are only available for some.
Positions in the CMD, however, can be compromised by the effects of
unresolved multiplicity on colours and magnitudes, as well as poor distances.

% COLOR-MAG BINARITY
Some of the systems are known to be binaries which may have been
resolved in one photometric band, but not in the catalogue providing the
$BV$ photometry.  There are three unresolved binary systems in our
sample of A-, F- and G-type members: HR 6749+HR 6750 (A5V+A5V), HD
199143AB (F7V+M2V) and HD 155555AB (G5V+K0V), all of which have
separations of less than $2''$. Plotting the combined photometry for
each binary would bias the ages inferred from the CMD and so we
deconstruct the combined photometric measurements into individual
component measurements. We calculate the individual component colours
and magnitudes for these systems following the technique of
\citet{Mermilliod92}, using the unresolved $V$-band magnitude, $B-V$
colour, and $\Delta V$-band magnitude between the two components.

% DISTANCES
Of the stars in our sample, both HIP 10679 and HIP 10680 (which
together form a G3V+F7V binary with a separation of $14''$) have
sufficiently large uncertainties in the measured parallaxes
\citep[$\varpi$ = 36.58\,$\pm$\,5.83 mas and 28.97\,$\pm$\,2.88 mas
  for HIP 10679 and 10680, respectively;][]{vanLeeuwen07} to warrant
calculating kinematic parallaxes. We therefore adopt the mean of the
UCAC4 proper motions for both components, and use the convergent point
solution derived in Section~\ref{revised_kinematic_parameters_bpmg} to
calculate a revised kinematic parallax for both HIP 10679 and HIP 10680
of $\varpi$ = 26.58\,$\pm$\,1.93 mas. We adopt this revised kinematic
parallax as opposed to the distance based on the revised
\emph{Hipparcos} reduction .

To derive the necessary bolometric corrections and
colour-$T_{\rm{eff}}$ relations to transform the model isochrones from
H-R diagram to CMD space we followed the formalism of \cite{Girardi02}
and calculated synthetic bolometric corrections BC$_{R_{\lambda}}$ in
a given bandpass with response function $R_{\lambda}$ as

\begin{eqnarray}
\label{bc_final}
  \mathrm{BC}_{_{R_{\lambda}}} & = & \mathrm{M}_{\rm{bol},\odot} -
  2.5\, \mathrm{log}\left(\frac{4\pi(10\mathrm{pc})^{2}F_{\rm{bol}}}{\mathrm{L}_{\odot}}\right) \\
  & + &~2.5\, \mathrm{log}\left(\frac{\int_{\lambda}\lambda
      F_{\lambda}10^{-0.4 A_{\lambda}}R_{\lambda}\, d\lambda}{\int_{\lambda}\lambda
    f^\circ_{\lambda}R_{\lambda}\, d\lambda}\right) -
   m^\circ_{_{R_{\lambda}}}. \nonumber
\end{eqnarray}

\noindent Here, $F_{\rm{bol}}=\sigma T_{\rm{eff}}^{4}$ is the total
flux emergent at the stellar surface, $F_{\lambda}$ is the flux at the
stellar surface and $A_{\lambda}$ is the extinction in the
bandpass. We note that all the BPMG members in this study lie within a
distance of $75\,\rm{pc}$ and are therefore subject to negligible
reddening \citep[see discussion in ][]{Pecaut13}, hence we adopt
$A_{\lambda}=0$. For the response functions $R_{\lambda}$ we adopt the
revised $BV$ bandpasses from \cite{Bessell12}. The term
$f^{\circ}_{\lambda}$ denotes a reference flux which produces a known
apparent magnitude $m^{\circ}_{\lambda}$. To ensure that the model
isochrones are as close to the standard Johnson $BV$ system as
possible, we adopt the CALSPEC
alpha\_lyr\_stis\_005\footnote{\url{http://www.stsci.edu/hst/observatory/cdbs/calspec.html}}
Vega spectrum as well as the zero-point (and additional zero-point)
offsets as described in \cite{Bessell12}. Finally, we adopt the
revised solar parameters from \citet{Mamajek12Fom} which take into
account recent total solar irradiance measurements from
\citet{Kopp11}: M$_{\rm{bol}, \odot}=4.755\,\rm{mag}$ and
L$_{\odot}=3.827 \times 10^{33}\,\rm{erg\,s^{-1}}$.

Fig.~\ref{fig:beta_pic} shows the $M_{V}, B-V$ CMDs for the A-, F- and
G-type BPMG members with the Yonsei-Yale (Y$^{2}$;
\citealp{Demarque04}), Dartmouth \citep{Dotter08}, Pisa
\citep*{Tognelli11} and PARSEC \citep{Bressan12} model isochrones
overlaid. The isochrones have been transformed into the observational
plane using the \textsc{atlas9} atmospheric models of
\citet{Castelli04}, for which we assume no $\alpha$-element
enhancement.

Despite the differences in the underlying assumptions adopted in the
evolutionary models (e.g. protosolar composition,
convection, etc.), it is clear that the ``classic'' group age of
$12\,\rm{Myr}$ is not supported by comparison of the CMD positions to
any of the model isochrones.  The A-type stars (HR 6070, HR 7329,
$\beta$ Pic, HIP 92024, HR 6749, and HR 6750) and the F0 star 51 Eri
are on the ZAMS (indeed, 51 Eri essentially defines the main-sequence
``turn-on"). All four sets of model isochrones do an admirable job of
tracing the ZAMS for the A1-F0 stars, i.e. the stars between
$\sim1.6-2.1\,\rm{M_{\odot}}$ have passed the final luminosity and
radius minimum of their contraction, and are now stably CNO-burning on
the main-sequence.  If the group were $12\,\rm{Myr}$ old, all four
sets of model isochrones would predict that only the two hottest
A0/A1-type stars (HR 6070 and HR 7329) would be on or near the ZAMS,
whereas all the cooler stars (including $\beta$ Pic itself!) should be
pre-MS, however this is clearly not the case.

The F3V star HR 9 ($B-V=0.39\,\rm{mag}$, $M_{V}=3.20\,\rm{mag}$)
appears to be in a region of the CMD between where the model
isochrones predict a second (penultimate) luminosity minima and the
final luminosity minimum corresponding to the ZAMS (where the first
luminosity minima occurs at the transition between the star being
mostly convective [Hayashi phase] and mostly radiative [Henyey
  phase]).  Evolutionary models predict that stars more massive than
about 1.2\,M$_{\odot}$ first contract to a
$^{12}$C(p,$\gamma$)-burning sequence -- essentially a ``{\it false
  ZAMS}'' -- before reaching the full CNO-burning ZAMS.  After the
core $^{12}$C is depleted to equilibrium levels with $^{14}$N via the
CN (Bethe-Weizs\"{a}cker) cycle, the core contracts further and the
temperature increases until the rest of the CNO-I cycle can complete
and $^{12}$C transitions from a consumed fuel to a catalyst, and
$^{12}$C returns to equilibrium levels [via
  $^{14}$N(p,$\gamma$)$^{15}$O($\beta^{+}\nu$)$^{15}$N and
  (p,$\alpha$)$^{12}$C reactions] \citep{Iben65, Clayton83}.  Using
the example of the star HR 9 (a $\sim$1.5\,M$_{\odot}$ star), the MESA
stellar evolution code \citep{Paxton11} predicts that a star of this
mass with protosolar initial composition takes $\sim$13\,Myr of pre-MS
evolution to contract to the penultimate luminosity minimum (``false
ZAMS''), and another $\sim$10\,Myr to reach the final CNO-burning ZAMS
at an age of $\sim$23\,Myr. Although the luminosity minima differ by
only $\sim 0.04\,\rm{dex}$ in log($L/\rm{L}_{\odot}$), the evolution
from the ``false ZAMS'' (penultimate pre-MS luminosity minimum) to the
ZAMS (final luminosity minimum) takes nearly as long as the pre-MS
contraction phase to the ``false ZAMS''. This stage is often ignored
in the literature on the evolution of young stars, but it can
correspond to a sizeable fraction of the early lifespan of a young
massive ($>$1.2 M$_{\odot}$) star.
% We dub this the ``Iben phase'',
% honouring the astrophysicist who appears to be the first to describe
% this epoch \citep{Iben65}. REMOVED AT EDITOR'S REQUEST 9/8/2014
HR 9 may hence be an example of a star in this phase where incomplete
nuclear CNO cycle burning has initiated, however core temperatures
have not yet reached the point where the full CNO cycle operates
catalytically and halts the star's contraction. The stars cooler than
HR 9 (type F4 and later) are overluminous compared to both the ZAMS
and ``false ZAMS", and seem to be pre-MS stars (see
Fig.~\ref{fig:beta_pic}; exceptions are HIP 10679 and HIP 10680 which
have particularly large parallax errors).

\begin{table*}
\caption[]{Median age estimates for the pre-MS F- and G-type BPMG
  members using both the ``classic" membership
  from \protect\cite{Zuckerman04} and the
  combined sample which includes additional members from
  \protect\cite{Malo13}.}  \centering
\begin{tabular}{c c c c c}
\hline
\multirow{2}{*}{BPMG Sample}& \multicolumn{4}{c}{Median age (Myr)}\\
& Yonsei-Yale & Dartmouth & Pisa & PARSEC\\
\hline
\cite{Zuckerman04} & $20\pm3$ & $21\pm3$ & $20\pm3$ & $21\pm2$\\
\cite{Zuckerman04} & \multirow{2}{*}{$21\pm4$} & \multirow{2}{*}{$22\pm3$} & \multirow{2}{*}{$22\pm2$} & \multirow{2}{*}{$21\pm3$}\\
+ \cite{Malo13} & & & &\\
\hline
F- and G-type isochronal age & \multicolumn{4}{c}{22\,Myr ($\pm 3\,\rm{Myr}$ statistical, $\pm 1\,\rm{Myr}$ systematic)}\\ 
\hline
\end{tabular}
\label{tab:ages}
\end{table*}

We can calculate an isochronal age for the BPMG using the F- and
G-type stars which are still in the pre-MS phase and estimate an age
for each star by linearly interpolating between finely-spaced grids of
model isochrones. Table~\ref{tab:ages} shows the median age estimates
for the BPMG from the four sets of model isochrones adopted in this
study. We derive ages (rounded to the nearest $1\,\rm{Myr}$) for both
(1) the ``classic" sample of members as defined by \citet{Zuckerman04}
and (2) the combined sample which includes additional members from
\citet{Malo13}, however we note that both ages are statistically
indistinguishable. We exclude HIP 10679 and HIP 10680 from the
analysis due to their large absolute magnitude uncertainties and the
fact that their CMD positions near or below the ZAMS preclude a useful
assessment of their isochronal ages (see
Fig.~\ref{fig:beta_pic}). Using the isochrones, we estimate the BPMG
age to be $22 \pm 3\,\rm{Myr}$, where the uncertainty is dominated by
$\pm 3\,\rm{Myr}$ statistical uncertainty, with a minor systematic
uncertainty of $\pm 1\,\rm{Myr}$ reflecting minor differences between
the isochrones.

\subsection{Comparison to other age estimates and adoption of a
final age}

While our new isochronal age for the F- and G-type pre-MS members of
BPMG ($22\pm 3\,\rm{Myr}$) is nearly twice as old as the classic BPMG
age of $\sim12\,\rm{Myr}$, it compares favourably with the recent LDB
age estimates by \citet{Binks14} and \citet{Malo14b} ($21 \pm
4\,\rm{Myr}$ and $26 \pm 3\,\rm{Myr}$, respectively). It is also
commensurate with the revised isochronal age from \citet{Malo14b}
($15-28\,\rm{Myr}$) for the K- and M-type pre-MS members using
evolutionary tracks which include a magnetic field model. Our new
isochronal age estimate is also in line with other recent estimates
based on Li depletion ($21 \pm 9\,\rm{Myr}$; \citealp{Mentuch08}) and
more recent kinematic estimates \citep{Torres06, Makarov07}. Indeed,
{\it none} of the studies over the past decade using a variety of
methods has estimated an age as low as the kinematic ages of $\sim
11-12\,\rm{Myr}$ reported by \citet{Ortega02}, \citet{Song03}, and
\citet{Ortega04}. As we concluded previously, it is not clear that a
unique kinematic age with small uncertainties can be quoted. However,
taking into account the recent LDB analyses by \citet{Binks14} and
\citet{Malo14b}, as well as the recent isochronal age analyses for the
K and M stars by \citet{Malo14b} and the F and G stars by this study, a
median BPMG age of $23 \pm 3\,\rm{Myr}$ ($\pm 2\,\rm{Myr}$
statistical, $\pm 2\,\rm{Myr}$ systematic) seems to adequately fit the
positions of the stars in both the H-R diagram and CMD as well as the
Li depletion pattern of the A- through M-type group members.

Whilst we have used contemporary sets of pre-MS evolutionary models to
derive an isochronal age, these models still neglect certain physical
phenomena such as the effects of stellar rotation and magnetic
fields. Recently, \cite{Malo14b} used the Dartmouth magnetic stellar
evolutionary models to derive an average isochronal age of between 15
and $28\,\rm{Myr}$ for the BPMG K- and M-type stars as well as an LDB
age of $26\pm3\,\rm{Myr}$. This study reported significant differences
in the luminosity (at a given effective temperature) between models
which include a magnetic field and those that do not, in the sense
that models with appreciable magnetic field strengths are more
luminous (e.g. $\Delta L_{\rm{bol}} \simeq 0.3\,\rm{dex}$ for
$B_{\rm{surf}}=2.5\,\rm{kG}$).  Unfortunately, these magnetic
evolutionary models are currently unavailable for the spectral types
we are concerned with in this study, however work on a full grid is
underway (Fieden, priv. comm.). Thus, at present we are unable to
provide a quantitative assessment of the effects of such models on an
isochronal age based on the F- and G-type BPMG members, although a
qualitative discussion is possible.

Literature measurements for surface magnetic fields on pre-MS F- and
G-type stars are extremely sparse and suggest field strengths of less
than $1\,\rm{kG}$ (see e.g. \citealp{Reiners12}).  Fig.~4 of
\cite{Malo14b} compares the position in the H-R diagram of magnetic and
non-magnetic evolutionary models, and from this it appears as though
there is a maximum difference of $\Delta L_{\rm{bol}} \lesssim
0.1\,\rm{dex}$ for a $1\,\rm{kG}$ model at effective temperatures less
than $3800\,\rm{K}$. It is unclear whether such a difference in
luminosity remains constant through the G- and into the F-type regime
(which seems unlikely given the large difference in the depth of the
convective zone across such spectral types), however if we assume a
conservative and constant upper limit of $0.1\,\rm{dex}$ then this
equates to $\simeq 0.25\,\rm{mag}$ difference in the absolute
bolometric magnitude. Such a difference in the absolute magnitude
would imply that the BPMG has an age of $\simeq 30\,\rm{Myr}$ based on
the contraction of the F- and G-type members. Given that this
``possible" age is consistent with the upper limit of the
\cite{Malo14b} age, it will therefore be interesting to see whether,
when the grid of magnetic evolutionary models is complete, consistency
between the isochronal ages of the F-, G-, K- and M-type stars (in
addition to the LDB age) is found. Furthermore, given that the adopted
magnetic field strength in the evolutionary models can have a
significant effect on the model isochrone -- and hence isochronal age
-- it is crucial to have some constraints on which value should be
applied to the models and over which spectral type range. Hence
calculating the field strength for a number of BPMG members would be
extremely beneficial for constraining what is essentially a free parameter.

\section{Conclusions}

Analysis of the kinematics of the $\beta$ Pictoris moving group
(BPMG), as well as isochronal age constraints based on the
main-sequence ``turn-on'' and pre-main-sequence (pre-MS) F- and G-type
stars, yields the following findings:

\begin{itemize}

\item {\it The trends in Galactic velocity versus position for the
  BPMG members show only marginal evidence for expansion with the best
  available velocity data.} The most significant velocity trend
  detected is velocity trend of $V$ as a function of $Y$ ($dV/dY$ =
  0.052\,$\pm$\,0.019 km\,s$^{-1}$\,pc$^{-1}$), which is only
  signficant at 2.7$\sigma$. Combining the expansion rates in $X$ and
  $Y$, one can derive a low accuracy expansion age of
  $21^{+10}_{-5}\,\rm{Myr}$ (1$\sigma$ uncertainty; however the
  2$\sigma$ range spans $13-59\,\rm{Myr}$).  Hence, while the
  velocity trends are at least consistent with an expansion age of
  $\sim 20\,\rm{Myr}$, the uncertainties remain large using revised
  \emph{Hipparcos} astrometry and modern radial velocity estimates.

\item {\it Integrating the orbits of the BPMG using a realistic
  Galactic potential, and tracing their orbits back in time shows that
  the size of the BPMG does not appear to have been significantly
  smaller in the past}. At 12 Myr ago, the dispersion in $X$ positions
  was smaller than the current value, but the dispersions in both the
  $Y$ and $Z$ positions were larger. {\it At this point we are not
    comfortable assigning a unique and useful traceback age for the
    BPMG.}
  
\item {\it The colour-magnitude diagram (CMD) positions for the A-, F-
  and G-type members of the BPMG appear to be consistent with an age
  almost twice as old as the classic age of 12\,Myr.} The appearance
  of the A0-F0 stars on/near the ZAMS (including $\beta$ Pic itself)
  is consistent with isochronal ages of $> 20\,\rm{Myr}$. The locus of
  members in the spectral type range F3-G9 are consistent with a
  pre-MS isochronal age of
  22\,$\pm\,3$\,(stat.)\,$\pm\,1$\,(sys.)\,\rm{Myr}, where the
  systematic uncertainty takes into account the scatter in inferred
  ages from examining four sets of published theoretical model
  isochrones.

\item {\it The CMD positions of the BPMG stars, as well as their
  pattern of Li depletion, appear to be consistent with a median
  consensus age of 23\,$\pm$\,3\,Myr.} This age is commensurate with
  (1) the appearance of the BPMG A-type members on the ZAMS (this
  study), (2) the CMD positions for the F- and G-type BPMG members
  compared to theoretical model isochrones (this study), (3) the
  isochronal ages for the pre-MS K- and M-type stars when including a
  treatment for magnetic fields \citep{Malo14b}, and (4) the Li
  depletion boundary \citep{Binks14,Malo14b}. The age is also
  consistent with other recent estimates based on Li depletion
  \citep{Mentuch08}, as well as kinematic analyses
  (\citealp{Torres06,Makarov07}), whose imprecise kinematic ages are at
  least consistent more with $\sim$20\,Myr than $\sim$12\,Myr. At this
  point, we are unable to reconcile these results with the older Li
  depletion age ($\sim 40\,\rm{Myr}$) estimated by \citet{Macdonald10}
  using magneto-convection models.

\end{itemize}

\citet{Song12} recently advocated adopting an expansion age estimate
of $12\,\rm{Myr}$ for the BPMG as a ``model-independent age'' for this
important young stellar group. Based on the results shown here we
would recommend against adopting {\it any} of the previously published
expansion ages for the BPMG, and disregard any conclusions regarding
the reliability of {\it other} age indicators based on comparison to
kinematic ages for the BPMG \citep[see also recent review by
][]{Soderblom13}. In particular, the conclusions by \citet{Song12}
that the mean ages of the Lower Centaurus-Crux (LCC) and Upper
Centaurus Lupus (UCL) subgroups of Sco-Cen are $\sim 10\,\rm{Myr}$ are
rendered invalid as their group ages were assessed through comparison
to kinematic ages for both the BPMG and the TW Hydrae association
(TWA). The Li depletion pattern for low-mass members of LCC and UCL
hint that it is intermediate in age between the BPMG and TWA.
However, mean isochronal ages for LCC and UCL of $16-17\,\rm{Myr}$
have been independently assessed for both the B-type main-sequence
``turn-off" members \citep{Mamajek02} and F-type pre-MS members
\citep{Pecaut12}.  The mean LCC and UCL ages estimated by
\citet{Mamajek02} and \citet{Pecaut12} are consistent with the Li
depletion pattern observed by \citet{Song12} if indeed BPMG is $\sim
23\,\rm{Myr}$ and TWA is $\sim 8-10\,\rm{Myr}$
\citep{Weinberger13,Ducourant14}.

\section*{Acknowledgements}

We thank Eric Jensen and Kevin Luhman for providing very detailed
comments and criticisms on an early draft of the paper, and Mark
Pecaut, Jackie Faherty, Rob Jeffries, Alexander Binks, Adric Reidel,
Joel Kastner, Lison Malo, and Val Rapson for discussions. We thank
Andrew Roberts for assistance with the NEMO dynamics code. Finally,
we thank the referee whose comments improved the manuscript. EEM
acknowledges support from NSF grant AST-1313029. EEM and CPMB
acknowledge support from the University of Rochester School of Arts
and Sciences.

\bibliographystyle{mn3e}
\bibliography{mamajek}

%\begin{thebibliography}{}
%\end{thebibliography}

\label{lastpage}

\end{document}